---

# Learning Mechanics and Game Mechanics Under the Perspective of Self-Determination Theory to Foster Motivation in Digital Game Based Learning.

**Jean-Nicolas Proulx[1], Margarida Romero[1], Sylvester Arnab[2]**

**Abstract**

Background: Using digital games for educational purposes has been associated with higher levels of **motivation** among learners of different educational levels. However, the underlying psychological factors involved in **digital game based learning** (**DGBL**) have been rarely analyzed considering **self-determination theory** (**SDT**, Ryan & Deci, 2000b); the relation of SDT with the **flow** experience (Csikszentmihalyi, 1990) has neither been evaluated in the context of DGBL.

Aim: This article evaluates DGBL under the perspective of SDT in order to improve the study of motivational factors in DGBL.

Results: In this paper, we introduce the **LMGM-SDT** theoretical framework, where the use of DGBL is analyzed through the **Learning Mechanics and Game Mechanics** mapping model (**LM-GM**, Arnab et al., 2015) and its relation with the components of the SDT. The implications for the use of DGBL in order to promote learners' motivation are also discussed.

[1] Université Laval, Québec, Canada

[2] Disruptive Media Learning Lab, Coventry University, UK

**Corresponding Author:**
Jean-Nicolas Proulx, Department of Teaching and Learning Studies, Université Laval, Québec, Canada.
Email : jean-nicolas.proulx.1@ulaval.ca

## Educational technologies and motivation

One of the persistent educational myths is that technology fosters learners' motivation by itself (Kleiman, 2000). Educational psychologists have issued warnings about the assumption that the use of new technologies in education, such as digital games, causes an increase in students' motivation (Amadieu & Tricot, 2014; De Bruyckere, Kirschner, & Hulshof, 2015). While in some studies, there seemed to be a link between the integration of technology in education and the learners' motivation (Papastergiou, 2009), in other research, that impact was not observed (e.g., Kebritchi, Hirumi, & Bai, 2010). Furthermore, Bitner and Bitner (2002) stressed the importance of the integration of technology into the curriculum as what matters the most when studying the technology effects on educational contexts. In other words, the technology itself does not necessarily foster student motivation (Kleiman, 2000), but the activities supported by technology could have the potential to have a positive impact on students' motivation (Price & Kadi-Hanifi, 2011). According to the literature review of Huang, Huang and Tschopp (2010, p. 790), "[motivation] is the essential element to initiate and sustain learning and performance". Hence, technology enhanced learning (TEL) activities need to be designed in a way that "provide adequate level of motivational stimuli" (Huang et al., 2010, p. 790). The study of the relation between technology and motivation is also present in the field of Digital Game Based Learning (DGBL) (Huang et al., 2010) and according to Whitton (2010, p. 596), it is also often assumed by researchers and practitioners that "computer games are intrinsically motivational for most, if not all, people". Outcomes pertaining to research in DGBL have led to inconsistent results related to the link between motivation and the use of digital games in education (Filsecker & Hickey, 2014). Furthermore, Kebritchi et al. (2010) discovered in their literature review on instructional games and motivation that only four of their 16 studies indicated that learners showed a motivational increase from using serious games. Also, according to Kebritchi et al. (2010, p. 428), "the studies mostly used mathematics games as treatments, achievement and motivation as dependent variables, and prior knowledge, computer experience, and language background as independent variables" (p. 428). In their literature review, instructional games were mostly used in mathematics and the student level varied from elementary to higher education.

We interpret the inconsistent results between DGBL and motivation according to two main factors. Firstly, DGBL integration in the classroom is a complex situation that might involve factors from multiple perspectives and levels (the learners' diversity, the teacher, the game and its mechanics, the DGBL integration in the learning situation, etc.). These diversity of factors, perspectives and levels of analysis result in a complex analysis of the effects of DGBL on motivation; multiple factors, which are not necessarily related to the digital game, such as teacher support and evaluation context, could mediate the relation between DGBL and the learners' motivation. Secondly, digital games in education come in great variety depending on their goals and inherent mechanics, which means that they can hardly be studied as a whole. In the same game, some mechanics could be associated with motivational effects for certain students in a certain context (e.g. competition as a motivational aspect among high confidence students), while other mechanics could even hinder motivation (e.g. collecting points could be perceived as boring). According to Squire (2011), good games have the potential to inspire interest, creativity, and social interaction. However, there are also games that, despite their intentions to be educational and playful, fail to engage certain types of learners.

In order to find out what makes a digital game achieve its educational and playfulness objectives, we make the hypothesis that we should consider the analysis of DGBL on a micro-level (considering the learning mechanics and game mechanics) rather than on a macro level (considering the game as a whole). The micro-level approach aims to study the DGBL according to the specific learning and games mechanics, which constitute the game. In order to achieve this aim, we define two specific research questions:

- Research question 1 (RQ1): What are the self-determination components related to the learners' motivation in a context of DGBL?
- Research question 2 (RQ2): How the self-determination components can be operationalized through the Learning Mechanics and Game Mechanics (LM-GM) model?

In order to answer these two research questions, the following sections provide a review of the literature that analyzes the DGBL effects on students' motivation at different levels of education. After reviewing these studies, we introduce the self-determination theory (SDT, Ryan & Deci, 2000b) as one of the most complete and actual motivation theories in the social sciences. We introduce the rationale of choosing the SDT for the analysis of DGBL. Finally, we analyze SDT components according to their links with the LM-GM model (Lim et al., 2013).

## Digital game based learning and motivation

The existing body of research that analyzed the use of digital games in education generally shows an overall positive link between the use of DGBL and their motivational effects among the learners (Groff, Howells, & Cranmer, 2010; Gros, 2007; Ritzko & Robinson, 2006). Over a review of more than 70 empirical research studies on the use of digital games in the classroom from Connolly, Boyle, MacArthur, Hainey and Boyle (2012), the most observed positive outcomes were *affectivity and motivation* (n = 33) and *learning* (n = 32). Also, according to Wouters, Van Nimwegen, Van Oostendorp and Van Der Spek (2013), the potential of games to foster intrinsic motivation is mentioned in much research. Players are willing to spend more time and energy to complete an activity when they see it as *fun* (Wouters et al., 2013). In addition, two of the key factors that are associated with digital games are autonomy (possibility to make relevant choices) and competence (the task represents a challenge while being achievable), which are core components of the self-determination theory that positively influence motivation (Wouters et al., 2013).

Aside from motivation, games also demonstrate the potential to inspire interest, creativity and social interactions from students (Squire, 2011). Although he sees great potential in the use of games in education, Squire (2005, p. 4) indicates that many contextual elements need to be taken into account for motivation to emerge: “motivation for the gamers in my study was thus not simply a ‘property’ or variable that they either had or did not have; motivation emerged through the intersection of students’ goals and life histories, the game's affordances, and the institutional context”. In his experiment, Squire (2005) attributes the appearance of motivation to a coherence between these many contextual elements and not only to the use of a game itself. Also, according to Filsecker and Hickey (2014, p. 138), “the kinds of cognitive strategies used might well be what distinguish motivated and unmotivated learners” (Renkl, 1997). Those inconsistencies lead us to conclude that not every game experience is perceived as fun by the players. Those inconsistencies also bring the hypothesis that digital games have indeed the potential to foster motivation and learning, but that this potential might not turn into a concrete form if certain requirements regarding the game, the learners and the learning context are not met.

Wouters et al. (2013) suggested that games that are considered as motivating by the players may lead to a greater engagement in the game play from them. This is why including a motivational theory, such as self-determination theory (SDT, Ryan & Deci, 2000b), in the game design and educational integration might increase the probabilities that the game would be considered ‘fun’ by the players. By examining digital games under the perspective of SDT, we might find clues on the different elements that games could include in their design in order to increase their motivational potential. Since we inspect the game design on a micro level, the LM-GM model from Lim et al. (2013) will be used, because it analyzes serious games according to their respective Learning Mechanics and Game Mechanics.

In the next sections, we will analyze the LM-GM model under the self-determination perspective in order to identify the self-determination components related to the learners’ motivation in a context of DGBL (RQ1) and to evaluate the self-determination components that can be operationalized through the Learning Mechanics and Game Mechanics (LM-GM) model (RQ2). Based on these analyses, we propose a theoretical framework allying LM-GM and SDT that could help to explore the underlying components that map the use of digital games to the motivational effects. The goal behind the combination of SDT and LM-GM would be to overcome the motivational limits of games and to contribute to a better analysis of the motivational factors of DGBL. In the next section, we will review the SDT as well as its main components.

## Self-determination theory (SDT) and motivation

### *Motivation theories in DGBL*

Two of the common motivational theories used with DGBL are the flow experience (e.g. Chen & Hwang, 2015; Franciosi, 2011; Garris, Ahlers, & Driskell, 2002; Kiili, Lainema, de Freitas,

& Arnab, 2014; Su & Hsaio, 2015) and the Cognitive Evaluation Theory with the use of intrinsic motivation (e.g. Erhel & Jamet, 2013; Huang et al., 2010; Yang, 2012).

The flow theory features a concept called the *flow experience* which was introduced by Csikszentmihalyi (1975). According to Csikszentmihalyi (1990), this concept can be defined as a complete state of cognitive absorption or engagement in a task, in which the individual is not affected by thoughts or emotions unrelated to the task. Kiili et al., (2014) extended the flow theory to include *playability* and demonstrated the potential of the flow theory in evaluating the quality of educational games.

The Cognitive Evaluation Theory from Ryan and Deci (2000a) categorizes motivation into two different types: intrinsic motivation and extrinsic motivation. "The concept of intrinsic motivation (IM) refers to behaviours performed out of interest and enjoyment. In contrast, extrinsic motivation (EM) pertains to behaviors carried out to attain contingent outcomes (Deci, 1971)" (Vallerand & Ratelle, 2002, p. 37).

Intrinsic motivation is often used to affirm that games in educational contexts are fun (e.g., Liu, Horton, Olmanson, & Toprac, 2011). Using a theory such as SDT instead of CET to evaluate motivation in games could be very beneficial since, as we can see in Ryan and Deci (2000b), SDT is a meta-theory that englobes CET. In the next sections, we introduce SDT and its advantages in relation to the CET model.

## *Self-determination theory (SDT)*

While SDT is different from the Cognitive Evaluation Theory and the Flow Theory, it does share common ground with both of them. Both Kowal & Fortier (1999) and Lee (2005) were able to establish a link between SDT and the flow theory. In Kowal and Fortier (1999), participants who had a self-determined motivation reached the highest states of flow experience. In Lee (2005), students with higher self-determined motivation were more likely to reach the flow experience and to be deeply engaged in their task.

SDT is also linked to Cognitive Evaluation Theory (CET), since SDT was created with the CET as its core. Deci and Ryan (2000) decided to combine CET with three other motivational theories in order to make a motivational theory that would be more complete. That new theory of motivation developed by psychologists Deci and Ryan (2000) was given the name Self-Determination Theory and is described as the following: "Self-determination theory focuses on the dialectic between the active, growth-oriented human organism and social contexts that either support or undermine people's attempts to master and integrate their experiences into a coherent sense of self" (Deci & Ryan, 2004, p. 27). In order to achieve growth, people should be able to develop the three following needs: autonomy, competence and relatedness.

**Autonomy**. This concept refers to the ability "to self-organize and regulate one's own behavior (and avoid heteronomous control), which includes the tendency to work toward inner coherence and integration among regulatory demands and goals" (Deci & Ryan, 2000, p. 252). In other words, it refers to "being the perceived origin or source of one's own behavior" (Deci & Ryan, 2004, p. 8).

**Connection or relatedness.** This concept refers to the need for belonging and attachment to other people as we can see in the next two definitions. “To seek attachments and experience feelings of security, belongingness, and intimacy with others” (Deci & Ryan, 2000, p. 252). “Relatedness refers to feeling connected to others, to caring for and being cared for by those others, to having a sense of belongingness both with other individuals and with one’s community (Beaumeister & Leary, 1995; Bowlby, 1979; Harlow, 1958; Ryan, 1995)” (Deci & Ryan, 2004, p. 7). Also, that need for relatedness concerns the sense of being with others and is not fueled by the attainment of specific outcomes such as social status gain (Deci & Ryan, 2004).

**Competence.** According to Deci and Ryan (2000, p. 252), *competence* means “to engage optimal challenges and experience mastery or effectance in the physical and social worlds”. Using White's (1959) definition of effectance, Klimmt & Hartmann (2006, p. 137) defined this concept as “the satisfaction of having imposed an effect on the environment”. Within SDT, the concept of competence means that individuals are seeking the pleasure of being effective with what they engage themselves (Deci & Ryan, 2000). That need for competence would lead people to “seek challenges that are optimal for their capacities and to persistently attempt to maintain and enhance those skills and capacities through activity” (Deci & Ryan, 2004, p. 7).

Those three components represent key elements that can positively influence intrinsic motivation as well as autonomous motivation (Deci & Ryan, 2000). We will further elaborate on the different types of motivation in the next section.

## *The continuum of autonomy and motivation within SDT*

Within SDT, different types of motivation are distinguished depending on the reasons or goals that are related to the emergence of a new action (Ryan & Deci, 2000b). According to this theory, “motivation concerns energy, direction, persistence and equifinality—all aspects of activation and intention” (Ryan & Deci, 2000b, p. 69). Then, motivation is divided into three categories (see Figure 1). Lack of motivation is referred to as *amotivation* (Ryan & Deci, 2009). Intrinsic motivation “refers to doing something because it is inherently interesting or enjoyable” (Ryan & Deci, 2000a, p. 55) and Extrinsic motivation “refers to doing something because it leads to a separable outcome” (Ryan & Deci, 2000a, p. 55).

In order for learners to feel self-determined within an educational context, we can assume that they would need to develop a certain level of competence, connections with other people (peers, teacher/s, parent/s…) within a certain margin of autonomy that would allow them to feel in control of the learning task and their activities within the task. To that, we can add a coherence between the task and students’ own goals.

## *SDT in education*

According to the literature review of Reeve (2002), the autonomy of learners is related to their well-being in the classroom. The literature review of Reeve (2002) also indicates that students benefit from having a teacher who allows them to have a certain level of autonomy in the classroom, which seems to allow students to develop their orientation to grow, feel more competent, more confident, more open-minded and creative.

| Amotivation | Extrinsic motivation | | | | Intrinsic Motivation |
|---|---|---|---|---|---|
| Non Regulation | External Regulation | Introjected Regulation | Identified Regulation | Integrated regulation | Intrinsic Regulation |
| Lack of motivation | Controlled Motivation | | Autonomous Motivation | | |
| Lowest Relative Autonomy | ←――――――――――――→ | | | | Highest Relative Autonomy |

**Table 1.** Continuum in the self-determination according to the type of regulation and autonomy (Ryan & Deci, 2009)

Almost a decade after proposing the SDT, Ryan and Deci (2009, p. 179) highlighted a positive link between teachers that facilitate a certain degree of autonomy and the degree of engagement and motivation felt by learners: "various studies of elementary and high school students (e.g., Hardre & Reeve, 2003; Jang, Reeve & Deci, 2007; Skinner & Belmont, 1993) have shown that teachers' autonomy support is related to students' autonomous motivation and engagement". So, these results are showing the possibility that students' perception of autonomy can be influenced by how 'autonomy supportive' their teacher is.

A literature review was performed by Taylor et al. (2014) on 18 studies that "assessed the relation of motivation types according to SDT to school achievement" (Taylor et al., 2014, p. 344) by using the Academic Motivation Scale (AMS) from Vallerand and Bissonnette (1992). That literature review pointed out a relation between STD and academic achievement. In secondary education, Taylor et al. (2014, p. 345) found out that "intrinsic motivation and identified regulation have a moderately strong, positive relation with school achievement". Also, they observed that "introjected and external regulation had a weaker, but significant negative relation with school achievement" (Taylor et al., 2014, p. 345) and that "amotivation had a strong, negative relation to school achievement" (Taylor et al., 2014, p. 345).

## *SDT within digital games*

According to Katz and Assor (2007, p. 439), a game is more likely to be considered 'fun' by students if they can relate to their personal life with the game or if they are offered meaningful choices that take into consideration their "needs, interests, goals, ability and cultural background". That led us to believe that we may be able to increase the odds that a game is considered motivating by integrating certain components into the game mechanics. According to SDT, those elements would be autonomy, competence, relatedness and a coherence between game's goals and students'.

In a context where we are working to foster very specific components within the mechanics of the game, a model such as the LM-GM can come in very handy. A model such as LM-GM that looks at a serious game design at a micro level becomes useful, since it gives the possibility to use game mechanics in a way that fosters SDT components.

## The learning mechanics game mechanics model and game design

In this section, we describe the LM-GM model from Lim et al. (2013) and how it can be used in game design. After that, we analyze its impact on games on which it has been used so far.

### *Introduction to LM-GM model*

There is a lack of robust methodology that informs the correlation between mechanics related to learning and playing, which introduces a huge challenge in guaranteeing efficacy and success (Arnab et al., 2015; Arnab & Clarke, 2016; Garris et al., 2002; Wilson et al., 2009). Existing frameworks for designing serious games "do not specifically target the analysis of the relationships between game mechanics and learning constructs, which is a key factor in the game design for learning" (Arnab et al., 2015, p. 394).

To map pedagogical constructs to entertaining gameplay, Lim et al. (2013) proposed the LM-GM and was evaluated by Arnab et al. (2015) as a Serious Game (SG) analysis guideline with positive outcomes. As a general overview, the pedagogical elements are viewed as an abstract interface while game elements are deemed as a concrete interface of SGs. This means that pedagogy and its methods are abstract (theoretical and conceptual), while game mechanics are concrete, i.e. by rules or algorithms.

Figure 2 lists an example of the learning mechanics (LMs) and the game mechanics (GMs). The mechanics that support both learning and games are regarded as abstract or concrete, depending on the mode of operation, where the abstract elements are more conceptual and the concrete counterparts are more explicit and implementable. The concrete mechanisms enable abstract concepts to work. The model is, however, descriptive and not prescriptive, "in the sense that it allows its users to freely relate learning and gaming mechanics to describe SG situations by drawing a map and filling a table" (Arnab et al., 2015, p. 396).

### *What we know so far of the impact of games designed using the LM-GM model in education*

The LM-GM model can be used to either aid SG design or game analysis, where it provides a concise means to map how ludic elements link to pedagogic intent directly based on a player's actions and game play, i.e. SG mechanics.

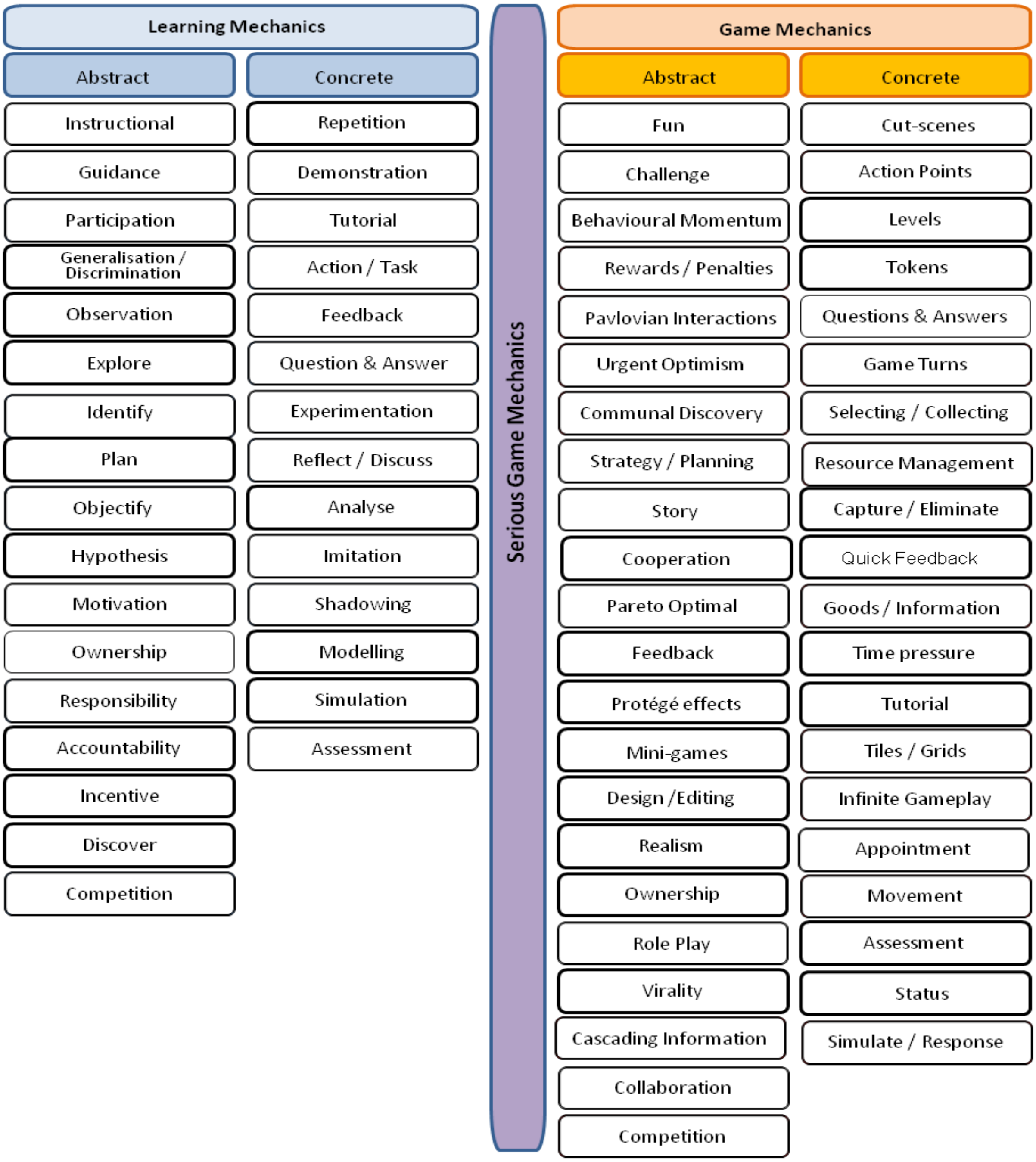

**Figure 1.** The LM-GM mapping framework (Arnab et al., 2015)

For instance, the model has been used to develop the design of the PR:EPARe game – a game to aid discourse on relationships and sexual pressure and coercion within a classroom setting (Arnab et al., 2013), where the abstract and concrete aspects of learning mechanics and game mechanics were used to identify the pedagogic delivery style that the game should take. Efficacy testing of the game solution was validated through a cluster-randomized controlled trial in local schools in the UK (n = 505), which indicated positive outcomes in favour of the game-based approach. Those results are based on self-reported measures of psychosocial preparedness for avoiding coercion, as well as the observation data on students and teachers experience through the process.

This evaluation of PR:EPARe has also led to the development of a more transdisciplinary methodology for game-based intervention design, where the different considerations taken to develop the game were infused (Arnab & Clarke, 2016). The LM-GM model is merged with the Mechanics-Dynamics-Aesthetics (MDA, Hunicke, LeBlanc, & Zubek, 2004) model, commonly used for entertainment games in order to marry the pedagogical (serious) aspects with the entertainment attributes of gameplay.

The LM-GM model has also been included as an important component towards informing the design of stealth assessment via a serious game within an intelligent scaffolding infrastructure (Baalsrud-Hauge et al., 2015), demonstrating the potential impact of LM-GM to inform stealth learning analytics via game mechanics. Reusability is, however, a key challenge due to the fact that serious games can be bespoke and not immediately re-purposable (Arnab & Clarke, 2016). Towards addressing this issue, Baalsrud-Hauge et al. (2015) analysed the mapping of the learning and game mechanics of two serious games – PR:EPARe and Seconds, which led to the definition of key SGM reuse principles, which demonstrated how the LM-GM model can be used to inform the analysis of serious game mechanics (SGM) towards identifying reusability.

## The game and learning mechanics under the perspective of the Self-determination theory (SDT)

Game mechanics involve a certain degree of interaction with the game artifact or the other players that are involved in the game activity. As a mechanism of interaction, 'retroactivity' represents a key aspect of the game mechanics that allows players to move forward within the game (Koster, 2013). Interactions with a game involve some degree of extrinsic motivation based on its retroactions. This characteristic has been highlighted by Dunwell, de Freitas and Jarvis (2011) as a positive aspect of games. A deep engagement of the player-learners inside the game universe can be interpreted as a sign that digital games used can also promote the emergence of autonomous motivation. The game and learning mechanics analysis under the perspective of the Self-determination theory (SDT) displayed in Figure 3, was developed with the intention of getting a bigger picture of the LMGM-SDT framework and advance the study of motivational factors in DGBL. The framework was created by mixing together the LM-GM classification based on Bloom's taxonomy from Arnab et al. (2015) and the self-determination continuum from Ryan and Deci (2009). The rationale behind that framework was to identify and analyze the game mechanics (intending to support the game experience) and learning mechanics (intending to support the learning activity) with a potential to support motivation. In order to analyse the LM-GM under the lens of SDT, we identified the learning mechanics and game mechanics that seemed to have a clearer link with the SDT components of the self-determination continuum from Ryan and Deci (2009), which ranges from amotivation, to extrinsic motivation and ends with intrinsic motivation as its highest level (represented in Figure 3 in the horizontal axis). In the vertical axis, Figure 3 is organized on the Bloom's taxonomy (last column in bold), from retention at the lowest level and creating in the higher level. The central columns display the game mechanics (left) and learning mechanics (right, grey color and italic) according to their extrinsic and intrinsic motivation potential.

Since games always provide a certain level of interaction with the game artifact or the other players (Koster, 2013), none of the game nor learning mechanics were placed into the amotivation zone of the table. Also, the learning outcomes and motivational potential will depend on the implementation of the mechanics within the learning situation where the game is integrated. This means that the same mechanic can be placed in different places depending on the concrete form it takes within the game environment. By classifying them both on self-determination continuum and on Bloom's ordered thinking skills, it becomes possible to evaluate the pedagogical outcome as well as the motivational potential of the mechanics that are used within the game.

Due to the neutral aspect or their motivational potential, some game and learning mechanics of the LM-GM model, such as the meta-game mechanic and realism mechanic, cannot be classified. The LM-GM model was based on Bloom's learning theory in Arnab et al. (2015) in order to introduce a taxonomy to classify the different Learning Mechanics and Game Mechanics. That way, game mechanics and learning mechanics can be used in a way that fosters very specific thinking skills and to make sure that the game as a whole remains an efficient learning tool. The same principle can also apply to motivation by exploiting game mechanics and learning mechanics in a way that fosters SDT components. In addition, considerable attention should be paid to the context and manner in which the different mechanics are integrated within the game, since it can affect the SDT continuum perception of the learner. For instance, the game mechanic 'progression' can be in both the second and fourth columns, because depending on how this mechanic is used, it can be perceived by the student as a controlling mechanic (e.g. If every student has to go through A, B, C, D and E in that specific order to reach F) or an autonomy supporting mechanic (e.g. If students are suggested a certain pathway by the game in order to reach F, but they are the one deciding how they want to reach F).

## Discussion

In this paper, we introduce a framework that combines both a game design model and a motivational theory. That framework aims to improve the motivational potential of games by allowing learners to develop their autonomous motivation.

Our first research question (RQ1) aimed to identify the self-determination components related to learners' motivation in a context of DGBL. In order to develop the learners' motivation, specifically autonomous motivation, which has been related to a positive impact on academic achievement (Taylor et al., 2014) and engagement (Lee, 2005), the game based learning activity should be in coherence with four requirements of the player activity: the students' goals (requirement 1), students' feeling of autonomy (requirement 2), competence (requirement 3) and relatedness (requirement 4) within the game. According to SDT, if these four requirements are met, then the game should foster autonomous motivation.

| Amotivation | Extrinsic motivation | | Intrinsic Motivation | | Thinking skills level |
|---|---|---|---|---|---|
| Non Regulation<br>Lack of motivation<br>Lowest Autonomy | External regulation -» Integrated regulation<br>Controlled Motivation | | Intrinsic Regulation<br>Autonomous Motivation<br>Highest Relative Autonomy | | |
| There are no learning mechanics neither game mechanics which aims to develop amotivation. | | *Accountability* | Designing<br>Ownership<br>Status/Titles<br>Strategy/planning | *Ownership*<br>*Planning*<br>*Responsibility* | **Creating** |
| | Action points<br>Assessment<br>Rewards/penalties | *Assessment*<br>*Incentive*<br>*External motivation* | Collaboration<br>Communal discovery<br>Game turns | *Collaboration*<br>*Hypothesis*<br>*Reflect/Discuss* | **Evaluating** |
| | Feedback | *Feedback*<br>*Shadowing* | | *Analyze*<br>*Experimentation*<br>*Identity* | **Analyzing** |
| | Progression (if too directed)<br>Time pressure | *Demonstration*<br>*Imitation*<br>*Simulation* | Competition<br>Cooperation<br>Movement<br>Progression<br>Selection/Collection<br>Simulate response | *Action/Task*<br>*Competition*<br>*Cooperation* | **Applying** |
| | Appointment<br>Cascading information<br>Questions and Tutorial | *Objectivity*<br>*Questions and answers*<br>*Tutorial* | Role-play | *Participation* | **Understanding** |
| | Behavioural Momentum<br>Cut-scenes<br>Story<br>Goods/Information<br>Pavlovian Interaction<br>Tokens | *Guidance*<br>*Instruction*<br>*Repetition* | Virality | *Discover*<br>*Explore*<br>*Generalization* | **Retention** |

**Table 2**. Game mechanics (left) and learning mechanics (right, grey color and italic) according to their extrinsic and intrinsic motivation potential.

Our second research question (RQ2) aimed to find a way to operationalize SDT components through the Learning Mechanics and Game Mechanics (LM-GM) model. As we saw in Figure 3, the LM-GM model aims to facilitate the operationalization of the game mechanics in a way that can foster different motivation types. In order to foster SDT components, a special attention needs to be paid to the nature of the mechanics that are used (which mechanics), since different mechanics will have different impacts. Also, as we saw in the previous section with the game mechanic *progression*, the way that mechanics are operationalized within the game can also have an impact on how they are perceived by players.

Analyzing the LM-GM from the perspective of SDT shows the need to evaluate the learning mechanics and game mechanics in the context they are used. For instance, a game designed with a game mechanic (e.g. communal discovery) could foster the player-learners' autonomy and self-determination. If there is too much external regulation from the instructional support provided by the teacher, the learners' perception of autonomy may be reduced. On the other hand, a game with mechanics that only promotes externally controlled motivation could be integrated in a more challenging task with higher degrees of autonomy (e.g. meaningful choices) and team engagement. In all cases, we should consider having a balance between many factors such as the level of external and intrinsic motivation provided by the digital game, the learners' level of competence and their regulation and co-regulation capabilities to avoid under and over regulation situations (Romero & Lambropoulous, 2011), which could end up having a negative impact on the learners' motivation and engagement.

The framework we assembled combines both the self-determination motivational theory and the LM-GM framework. While both SDT and LM-GM seems individually promising, we have yet to explore their compatibility. In order to investigate the compatibility and the pedagogical relevance of the framework, we still need to experiment it in a concrete class setting. However, the present paper does bring forward evidences that a game design that meets the four requirements of the player activity should contribute to enjoyment, motivation and engagement.

## Acknowledgments

We would like to thank the reviewers and Nicola Whitton for their thorough reviews and, in particular, for their constructive critique and suggestions for improvement.

## Declaration of Conflicting Interests

The authors declared no potential conflicts of interest with respect to the research, authorship, and/or publication of this article.

## Funding

This research received no specific grant from any funding agency in the public, commercial, or not-for-profit sectors.

## References

Amadieu, F., & Tricot, A. (2014). *Apprendre avec le numérique: mythes et réalités*. Retz.

Arnab, S., Brown, K., Clarke, S., Dunwell, I., Lim, T., Suttie, N., … De Freitas, S. (2013). The development approach of a pedagogically-driven serious game to support Relationship and Sex Education (RSE) within a classroom setting. *Computers & Education*, *69*, 15–30. http://doi.org/10.1016/j.compedu.2013.06.013

Arnab, S., & Clarke, S. (2016). Towards a trans-disciplinary methodology for a game-based intervention development process. *British Journal of Educational Technology*. http://doi.org/10.1111/bjet.12377

Arnab, S., Lim, T., Carvalho, M. B., Bellotti, F., Freitas, S., Louchart, S., … De Gloria, A. (2015). Mapping learning and game mechanics for serious games analysis. *British Journal of Educational Technology*, *46*(2), 391–411. http://doi.org/10.1111/bjet.12113

Baalsrud-Hauge, J. M., Stanescu, I. A., Arnab, S., Ger, P. M., Lim, T., Serrano-Laguna, A., … others. (2015). Learning through analytics architecture to scaffold learning experience through technology-based methods. *International Journal of Serious Games*, *2*(1), 29–44. http://doi.org/10.17083/ijsg.v2i1.38

Bitner, N., & Bitner, J. (2002). Integrating technology into the classroom: Eight keys to success. *Journal of Technology and Teacher Education*, *10*(1), 95–100.

Chen, C.-H., & Hwang, G.-J. (2015). Improving learning achievements, motivations and flow with a progressive prompt-based mobile gaming approach. In *IIAI 4th International Congress on Advanced Applied Informatic* (pp. 297–302). Okayama: IEEE. http://doi.org/10.1109/IIAI-AAI.2015.163

Connolly, T. M., Boyle, E. A., MacArthur, E., Hainey, T., & Boyle, J. M. (2012). A systematic literature review of empirical evidence on computer games and serious games. *Computers & Education*, *59*(2), 661–686. http://doi.org/10.1016/j.compedu.2012.03.004

Csikszentmihalyi, M. (1975). Play and intrinsic rewards. *Journal of Humanistic Psychology*, *15*(3), 41–63.

Csikszentmihalyi, M. (1990). *Flow: The psychology of optimal experience*. New York: Cambridge University Press.

De Bruyckere, P., Kirschner, P. A., & Hulshof, C. D. (2015). *Urban myths about learning and education*. New York: Academic Press.

Deci, E. L., & Ryan, R. M. (2000). The "what" and "why" of goal pursuits: Human needs and the self-determination of behavior. *Psychological Inquiry*, *11*(4), 227–268.

Deci, E. L., & Ryan, R. M. (2004). Overview of self-determination theory: An organismic dialectical perspective. In E. L. Deci & R. M. Ryan (Eds.), *Handbook of self-determination research* (pp. 3–33). Rochester, NY: University Of Rochester Press.

Dunwell, I., De Freitas, S., & Jarvis, S. (2011). Four-dimensional consideration of feedback in serious games. In S. De Freitas & P. Maharg (Eds.), *Digital games and learning* (pp. 42–62). Continuum Publishing.

Erhel, S., & Jamet, E. (2013). Digital game-based learning: Impact of instructions and feedback on motivation and learning effectiveness. *Computers & Education*, *67*, 156–167.

Filsecker, M., & Hickey, D. T. (2014). A multilevel analysis of the effects of external rewards on elementary students' motivation, engagement and learning in an educational game. *Computers & Education*, *75*, 136–148. http://doi.org/10.1016/j.compedu.2014.02.008

Franciosi, S. J. (2011). A comparison of computer game and language-learning task design using flow theory. *CALL-EJ*, *12*(1), 11–25.

Garris, R., Ahlers, R., & Driskell, J. E. (2002). Games, motivation, and learning: A research and practice model. *Simulation & Gaming*, *33*(4), 441–467.

Groff, J., Howells, C., & Cranmer, S. (2010). *The impact of console games in the classroom: Evidence from schools in Scotland*. UK: Futurelab.

Gros, B. (2007). Digital games in education: The design of games-based learning environments. *Journal of Research on Technology in Education*, *40*(1), 23–38.

Hauge, J. B., Stanescu, I. A., Stefan, A., Carvalho, M. B., Lim, T., Louchart, S., & arnab, S. (2015). Serious Games Mechanics and oppotunities for reuse. In *eLearning & Software for Education* (pp. 19–27). Bucharest, Romania. http://doi.org/10.12753/2066-026X-15-000

Huang, W.-H., Huang, W.-Y., & Tschopp, J. (2010). Sustaining iterative game playing processes in DGBL: The relationship between motivational processing and outcome processing. *Computers & Education*, *55*(2), 789–797.

Hunicke, R., LeBlanc, M., & Zubek, R. (2004). MDA: A formal approach to game design and game research. Presented at the Game Developers Conference. Retrieved from http://www.cs.northwestern.edu/~hunicke/MDA.pdf

Katz, I., & Assor, A. (2007). When choice motivates and when it does not. *Educational Psychology Review*, *19*(4), 429–442.

Kebritchi, M., Hirumi, A., & Bai, H. (2010). The effects of modern mathematics computer games on mathematics achievement and class motivation. *Computers & Education*, *55*, 427–443. http://doi.org/10.1016/j.compedu.2010.02 .007

Kiili, K., Lainema, T., de Freitas, S., & Arnab, S. (2014). Flow framework for analyzing the quality of educational games. *Entertainment Computing*, *5*(4), 367–377. http://doi.org/10.1016/j.entcom.2014.08.002

Kleiman, G. M. (2000). Myths and realities about technology in K-12 schools. *The Online Journal of the Leadership and the New Technologies Community*, (14), 1–8.

Klimmt, C., & Hartmann, T. (2006). Effectance, self-efficacy, and the motivation to play video games. In P. Vorderer & J. Bryant (Eds.), *Playing video games: Motives, responses, and consequences* (pp. 133–145). Mahwah, NJ, US: Lawrence Erlbaum Associates Publishers.

Koster, R. (2013). *Theory of fun for game design* (2nd ed.). Sebastopol, CA: O'Reilly Media.

Kowal, J., & Fortier, M. S. (1999). Motivational determinants of flow: Contributions from self-determination theory. *The Journal of Social Psychology*, *139*(3), 355–368. http://doi.org/10.1080/00224549909598391

Lee, E. (2005). The Relationship of Motivation and Flow Experience to Academic Procrastination in University Students. *The Journal of Genetic Psychology*, *166*(1), 5–14. http://doi.org/10.3200/GNTP.166.1.5-15

Lim, T., Louchart, S., Suttie, N., Ritchie, J. M., Aylett, R. S., Stanescu, I. A., … Moreno-Ger, P. (2013). Strategies for effective digital games development and implementation. In Y. Baek & N. Whitton (Eds.), *Cases on digital game-based learning: Methods, models, and strategies* (pp. 168–198). Hershey, PA.

Liu, M., Horton, L., Olmanson, J., & Toprac, P. (2011). A study of learning and motivation in a new media enriched environment for middle school science. *Educational Technology Research and Development*, *59*(2), 249–265. http://doi.org/10.1007/s11423-011-9192-7

Papastergiou, M. (2009). Digital game-based learning in high school computer science education: Impact on educational effectiveness and student motivation. *Computers & Education*, *52*(1), 1–12.

Price, F., & Kadi-Hanifi, K. (2011). E-motivation! The role of popular technology in student motivation and retention. *Research in Post-Compulsory Education*, *16*(2), 173–187. http://doi.org/10.1080/13596748.2011.575278

Reeve, J. (2002). Self-determination theory applied to educational settings. In E. L. Deci & R. M. Ryan (Eds.), *Handbook of self-determination research* (pp. 183–203). Rochester, NY: University Of Rochester Press.

Renkl, A. (1997). Learning from worked-out examples: A study on individual differences. *Cognitive Science*, *21*(1), 1–29. http://doi.org/10.1207/s15516709cog2101_1

Ritzko, J. M., & Robinson, S. (2006). Using games to increase active learning. *Journal of College Teaching & Learning (TLC)*, *3*(6), 45–50. http://doi.org/10.19030/tlc.v3i6.1709

Romero, M., & Lambropoulous, N. (2011). Internal and external regulation to support knowledge construction and convergence in computer supported collaborative learning (CSCL). *Electronic Journal of Research in Educational Psychology*, *9*(1), 309–330.

Ryan, R. M., & Deci, E. L. (2000a). Intrinsic and extrinsic motivations: Classic definitions and new directions. *Contemporary Educational Psychology*, *25*(1), 54–67. http://doi.org/10.1006/ceps.1999.1020

Ryan, R. M., & Deci, E. L. (2000b). Self-determination theory and the facilitation of intrinsic motivation, social development, and well-being. *American Psychologist*, *55*(1), 68–78. http://doi.org/10.1037/0003-066X.55.1.68

Ryan, R. M., & Deci, E. L. (2009). Promoting self-determined school engagement : Motivation, learning and well-being. In K. R. Wentzel & A. Wigfield (Eds.), *Handbook of motivation at school* (pp. 171–196). New York, NY: Routledge.
Squire, K. (2005). Changing the game: What happens when video games enter the classroom. *Innovate: Journal of Online Education*, *1*(6). Retrieved from http://website.education.wisc.edu/~kdsquire/tenure-files/26-innovate.pdf
Squire, K. (2011). *Video Games and Learning: Teaching and Participatory Culture in the Digital Age.* New York, NY: Teacher College Press.
Su, C.-H., & Hsaio, K.-C. (2015). Developing and evaluating gamifying learning system by using flow-based model. *Eurasia Journal of Mathematics, Science & Technology Education*, *11*(6), 1283–1306.
Taylor, G., Jungert, T., Mageau, G. A., Schattke, K., Dedic, H., Rosenfield, S., & Koestner, R. (2014). A self-determination theory approach to predicting school achievement over time: the unique role of intrinsic motivation. *Contemporary Educational Psychology*, *39*(4), 342–358. http://doi.org/10.1016/j.cedpsych.2014.08.002
Vallerand, R. J., & Bissonnette, R. (1992). Intrinsic, extrinsic, and amotivational styles as predictors of behavior: A prospective study. *Journal of Personality*, *60*(3), 599–620. http://doi.org/10.1111/j.1467-6494.1992.tb00922.x
Vallerand, R. J., & Ratelle, C. F. (2002). Intrinsic and extrinsic motivation: A hierarchical model. In E. L. Deci & R. M. Ryan (Eds.), *Handbook of self-determination research* (Vol. 128, pp. 37–69). Rochester, NY: The University of Rochester Press.
White, R. W. (1959). Motivation reconsidered: the concept of competence. *Psychological Review*, *66*(5), 297–333. http://doi.org/10.1037/h0040934
Whitton, N. (2010). Game engagement theory and adult learning. *Simulation & Gaming*, *42*(5), 596–609. http://doi.org/http://dx.doi.org/10.1177/1046878110378587
Wilson, K. A., Bedwell, W. L., Lazzara, E. H., Salas, E., Burke, C. S., Estock, J. L., … Conkey, C. (2009). Relationships between game attributes and learning outcomes review and research proposals. *Simulation & Gaming*, *40*(2), 217–266. http://doi.org/10.1177/1046878108321866
Wouters, P., van Nimwegen, C., van Oostendorp, H., & van Der Spek, E. D. (2013). A meta-analysis of the cognitive and motivational effects of serious games. *Journal of Educational Psychology*, *105*(2), 249–265. http://doi.org/10.1037/a0031311
Yang, Y.-T. C. (2012). Building virtual cities, inspiring intelligent citizens: Digital games for developing students' problem solving and learning motivation. *Computers & Education*, *59*(2), 365–377. http://doi.org/10.1016/j.compedu.2012.01.012

## Author Biographies

**Jean-Nicolas Proulx** is a PhD student in Educational Technologies and an Auxiliary Teacher in the Bachelor's Degree in Education and Primary Education program (BEPEP) at Université Laval. His main research interests are serious games, academic motivation, and academic success with a special focus on Game Based Learning. He is currently doing his thesis on the usage of Game Based Learning in order to increase academic success and motivation in high school history class setting.

**Margarida Romero is** professor of educational technology at Université Laval (Canada). Her research is oriented towards the inclusive, humanistic and creative uses of technologies (creative design and programming of games and educational robots) for the development of the 21st century skills across the lifespan: cooperation and communication, problem solving, creativity and computational thinking. She wrote the children's' book "Vibot the robot" aiming to demystify programming and robotics for children, their parents and their teachers.

**Dr. Sylvester Arnab** is a Reader in Games Science at Coventry University, UK co-leading research at the Disruptive Media Learning Lab (DMLL). With more than 10 years research experience in simulation, serious games and gamification combined, his research interests include gameful, playful and pervasive designs that transform ordinary tasks into extraordinary experiences. As lead of the Game Science research at the DMLL, Sylvester is also leading the EU Horizon 2020 BEACONING project (worth €5.9 million with 15 partners), which aims to foster 'anytime anywhere' learning using pervasive, context-aware and gamified techniques.